\documentclass[twocolumn,showpacs,showkeys,amsmath,amssymb,superscriptaddress,floatfix,nofootinbib]{revtex4}

\usepackage{graphicx}
\usepackage{bm} 
\usepackage{subfigure}

\makeatletter
\newcommand\erfc{\mathop{\operator@font erfc}\nolimits}
\def\slashchar#1{\setbox0=\hbox{$#1$}
   \dimen0=\wd0 \setbox1=\hbox{/} \dimen1=\wd1
   \ifdim\dimen0>\dimen1 \rlap{\hbox to \dimen0{\hfil/\hfil}} #1
   \else  \rlap{\hbox to \dimen1{\hfil$#1$\hfil}} / \fi}

\begin{document}
 
\title{
Projection method for boost-invariant and cylindrically symmetric dissipative hydrodynamics
\footnote{This work was supported in part by the Polish Ministry
of Science and Higher Education under Grants No. N N202
263438 and No. N N202 288638.}}

\author{Wojciech Florkowski} 
\email{Wojciech.Florkowski@ifj.edu.pl}
\affiliation{Institute of Physics, Jan Kochanowski University, PL-25406~Kielce, Poland} 
\affiliation{The H. Niewodnicza\'nski Institute of Nuclear Physics, Polish Academy of Sciences, PL-31342 Krak\'ow, Poland}

\author{Radoslaw Ryblewski} 
\email{Radoslaw.Ryblewski@ifj.edu.pl}
\affiliation{The H. Niewodnicza\'nski Institute of Nuclear Physics, Polish Academy of Sciences, PL-31342 Krak\'ow, Poland}

\date{November 22, 2011}

\begin{abstract}
Tensors describing boost-invariant and cylindrically symmetric expansion of a relativistic dissipative fluid are decomposed in a suitable chosen basis of projection operators.  This leads to a simple set of scalar equations which determine the fluid behavior. As special examples, we discuss the case of the Israel-Stewart theory and the model of highly-anisotropic and strongly-dissipative hydrodynamics ADHYDRO. We also introduce the matching conditions between the ADHYDRO description suitable for the very early stages of heavy-ion collisions and the Israel-Stewart theory applicable for later stages when the system is close to equilibrium. 
\end{abstract}



\maketitle 

%%%%%%%%%%%%%%%%%%%%%%%%%%%%%%%%%%%%%%%%%%%%%%%%%%%%%%%%%%%%%%%%%%%%%%%%%%%%%%%%%%%%%%%%%%%%%%%%%%%%
\section{Introduction}
\label{sect:intro}
%%%%%%%%%%%%%%%%%%%%%%%%%%%%%%%%%%%%%%%%%%%%%%%%%%%%%%%%%%%%%%%%%%%%%%%%%%%%%%%%%%%%%%%%%%%%%%%%%%%%%

Soft-hadronic observables measured in the ultra-relativistic heavy-ion experiments may be very well described by the standard perfect-fluid hydrodynamics (for a recent review see \cite{Florkowski:2010zz}) or by dissipative hydrodynamics with a small viscosity to entropy ratio \cite{Chaudhuri:2006jd,Dusling:2007gi,Luzum:2008cw,Song:2007fn,Bozek:2009dw,Schenke:2010rr}. These approaches assume generally that the produced system reaches a state of local thermal equilibrium within a fraction of a fermi~\footnote{We use the natural system of units where $\hbar=c=1$. The metric tensor $g_{\mu \nu} = \hbox{diag}(1,-1,-1,-1)$.}. 

On the other hand, many microscopic approaches assume that the produced system is initially highly anisotropic in the momentum space, e.g., see \cite{Bjoraker:2000cf}. High anisotropies present at the early stages of relativistic heavy-ion collisions exclude formally the application of the perfect-fluid and dissipative hydrodynamics. This situation has triggered development of several approaches which combine the anisotropic early evolution with a later perfect-fluid \cite{Sinyukov:2006dw,Gyulassy:2007zz,Broniowski:2008qk,Ryblewski:2010tn} or viscous \cite{Martinez:2009ry,Bozek:2010aj} description. Very recently, a concise model describing consistently different stages of heavy-ion collisions  has been proposed in Refs. \cite{Florkowski:2010cf,Ryblewski:2010bs,Ryblewski:2010ch,Ryblewski:2011aq} (a highly-Anisotropic and strongly-Dissipative HYDROdynamics, ADHYDRO), see also a similar work that has been presented in Refs. \cite{Martinez:2010sc,Martinez:2010sd}.

In this paper, in order to analyze in more detail the connections between different effective descriptions of very early stages of heavy-ion collisions we consider a simplified case of the boost-invariant and cylindrically symmetric expansion of matter. In the first part of the paper, we introduce tensors that form a suitable basis for decompositions of various tensors characterizing dissipative fluids. Then, we use this basis to analyze the Israel-Stewart \cite{Israel:1979wp,Muronga:2003ta} and ADHYDRO equations \cite{Florkowski:2010cf,Ryblewski:2010bs,Ryblewski:2010ch,Ryblewski:2011aq}. Finally, we show how the solutions of the ADHYDRO model may be matched with the solutions of the Israel-Stewart theory. The last result may be treated as a generalization of the approach presented in \cite{Martinez:2009ry} where no transverse expansion of matter was considered.

The formal results presented in this paper, when implemented as numerical procedures, may be used to model the behavior of matter produced at the very early stages of heavy-ion collisions. Of course, the use of boost-invariance and cylindrical symmetry implies that this description should be limited at the moment to central collisions and the central rapidity region. A generalization of our framework to more complicated geometries is a work in progress. 

%%%%%%%%%%%%%%%%%%%%%%%%%%%%%%%%%%%%%%%%%%%%%%%%%%%%%%%%%%%%%%%%%%%%%%%%%%%%%%%%%%%%%%%%%%%%%%%%%%%%%
%%%%%%%%%%%%%%%%%%%%%%%%%%%%%%%%%%%%%%%%%%%%%%%%%%%%%%%%%%%%%%%%%%%%%%%%%%%%%%%%%%%%%%%%%%%%%%%%%%%%%
\section{Boost-invariant and cylindrically symmetric flow}
\label{sect:flow}
%%%%%%%%%%%%%%%%%%%%%%%%%%%%%%%%%%%%%%%%%%%%%%%%%%%%%%%%%%%%%%%%%%%%%%%%%%%%%%%%%%%%%%%%%%%%%%%%%%%%%
%%%%%%%%%%%%%%%%%%%%%%%%%%%%%%%%%%%%%%%%%%%%%%%%%%%%%%%%%%%%%%%%%%%%%%%%%%%%%%%%%%%%%%%%%%%%%%%%%%%%%

The space-time coordinates and the four-vector describing the hydrodynamic flow are denoted in the standard way as 
$x^\mu = \left( t, x, y, z \right)$ and 
\begin{equation}
U^\mu = \gamma (1, v_x, v_y, v_z), \quad \gamma = (1-v^2)^{-1/2}.
\label{Umu}
\end{equation}
For boost-invariant and cylindrically symmetric systems, we may use the following parametrization
\begin{eqnarray}
U^0 &=& \cosh \theta_\perp \cosh \eta_\parallel, \nonumber \\
U^1 &=& \sinh \theta_\perp \cos  \phi,           \nonumber \\
U^2 &=& \sinh \theta_\perp \sin  \phi,           \nonumber \\
U^3 &=& \cosh \theta_\perp \sinh \eta_\parallel,
\label{Umu}
\end{eqnarray}
where $\theta_\perp$ is the transverse fluid rapidity defined by the formula
\begin{equation}
v_\perp = \sqrt{v_x^2+v_y^2} = \tanh \theta_\perp,
\label{thetaperp}
\end{equation} 
$\eta_\parallel$ is the space-time rapidity,
\begin{eqnarray}
\eta_\parallel = \frac{1}{2} \ln \frac{t+z}{t-z}, 
\label{etapar} 
\end{eqnarray}
and $\phi$ is the azimuthal angle
\begin{equation}
\phi = \arctan \frac{y}{x}.
\label{phi}
\end{equation}

In addition to $U^\mu$ we define three other four-vectors. The first one, $Z^\mu$, defines the longitudinal direction that plays a special role due to the initial geometry of the collision, 
\begin{eqnarray}
Z^0 &=& \sinh \eta_\parallel, \nonumber \\
Z^1 &=& 0,           \nonumber \\
Z^2 &=& 0,           \nonumber \\
Z^3 &=& \cosh \eta_\parallel.
\label{Zmu}
\end{eqnarray}
The second four-vector, $X^\mu$, defines a transverse direction to the beam,
\begin{eqnarray}
X^0 &=& \sinh \theta_\perp \cosh \eta_\parallel, \nonumber \\
X^1 &=& \cosh \theta_\perp \cos  \phi,           \nonumber \\
X^2 &=& \cosh \theta_\perp \sin  \phi,           \nonumber \\
X^3 &=& \sinh \theta_\perp \sinh \eta_\parallel,
\label{Xmu}
\end{eqnarray}
while the third four-vector, $Y^\mu$, defines the second transverse direction~\footnote{By analogy to the terminology used in interferometry (HBT) studies, one may say that the four-vector $X^\mu$ defines the {\it out} direction, while $Y^\mu$ defines the {\it side} direction.},
\begin{eqnarray}
Y^0 &=& 0, \nonumber \\
Y^1 &=& -\sin  \phi,           \nonumber \\
Y^2 &=& \cos  \phi,           \nonumber \\
Y^3 &=& 0.
\label{Ymu}
\end{eqnarray}

The four-vector $U^\mu$ is time-like, while the four-vectors $Z^\mu, X^\mu, Y^\mu$ are space-like. In addition, they are all orthogonal to each other, 
\begin{eqnarray}
U^2 &=& 1, \quad Z^2 = X^2 = Y^2 = -1, \nonumber \\
U \cdot Z &=& 0, \quad U \cdot X = 0, \quad U \cdot Y = 0, \nonumber \\
Z \cdot X &=& 0, \quad Z \cdot Y = 0, \quad X \cdot Y = 0.
\label{norm}
\end{eqnarray}

All these properties are most easily seen in the {\it local rest frame} of the fluid element (LRF), where we have $\theta_\perp = \eta_\parallel = \phi = 0$ and
\begin{eqnarray}
U &=& (1,0,0,0),           \nonumber \\
Z &=& (0,0,0,1),           \nonumber \\
X &=& (0,1,0,0),           \nonumber \\
Y &=& (0,0,1,0).
\label{LRF}
\end{eqnarray}

In the formalism of dissipative hydrodynamics one uses the operator $ \Delta^{\mu \nu} = g^{\mu \nu} - U^\mu U^\nu$, that projects on the three-dimensional space orthogonal to $U^\mu$. It can be shown that ~\footnote{One may check easily that (\ref{Delta}) holds in LRF. Hence, as a tensor equation, (\ref{Delta}) should hold in all reference frames.}
\begin{equation}
\Delta^{\mu \nu} = -X^\mu X^\nu - Y^\mu Y^\nu - Z^\mu Z^\nu.
\label{Delta}
\end{equation}
Using Eqs. (\ref{norm}) we find that $Z^\mu, X^\mu$ and $Y^\mu$ are the eigenvectors of $\Delta^{\mu \nu}$,
\begin{equation}
\Delta^{\mu \nu} X_\nu = X^\mu, \quad  \Delta^{\mu \nu} Y_\nu = Y^\mu, \quad 
\Delta^{\mu \nu} Z_\nu = Z^\mu.
\label{eigen}
\end{equation}

%%%%%%%%%%%%%%%%%%%%%%%%%%%%%%%%%%%%%%%%%%%%%%%%%%%%%%%%%%%%%%%%%%%%%%%%%%%%%%%%%%%%%%%%%%%%%%%%%%%%%
%%%%%%%%%%%%%%%%%%%%%%%%%%%%%%%%%%%%%%%%%%%%%%%%%%%%%%%%%%%%%%%%%%%%%%%%%%%%%%%%%%%%%%%%%%%%%%%%%%%%%
\section{Expansion and shear tensors}
\label{sect:exp-shear}
%%%%%%%%%%%%%%%%%%%%%%%%%%%%%%%%%%%%%%%%%%%%%%%%%%%%%%%%%%%%%%%%%%%%%%%%%%%%%%%%%%%%%%%%%%%%%%%%%%%%%
%%%%%%%%%%%%%%%%%%%%%%%%%%%%%%%%%%%%%%%%%%%%%%%%%%%%%%%%%%%%%%%%%%%%%%%%%%%%%%%%%%%%%%%%%%%%%%%%%%%%%

In this Section we follow the standard definitions of the expansion and shear tensors \cite{Muronga:2003ta} and show that they can be conveniently decomposed in the basis of the tensors obtained as products of the four-vectors $X^\mu$, $Y^\mu$ and $Z^\mu$.

The {\it expansion} tensor $\theta_{\mu \nu}$ is defined by the expression
\begin{equation}
\theta_{\mu \nu} = \Delta^\alpha_\mu \Delta^\beta_\nu \partial_{(\beta} U_{\alpha)},
\label{theta-munu}
\end{equation}
where the brackets denote the symmetric part of $\partial_{\beta} U_{\alpha}$. Using Eqs. (\ref{Umu}) in the definition of the expansion tensor (\ref{theta-munu}) and also using Eqs. (\ref{Zmu})--(\ref{Ymu}),  we may verify that the following decomposition holds
\begin{equation}
\theta^{\mu \nu} = \theta_X X^\mu X^\nu + \theta_Y Y^\mu Y^\nu +  \theta_Z Z^\mu Z^\nu, 
\label{theta-dec}
\end{equation}
where
\begin{equation}
\theta_X = X_\mu X_\nu \theta^{\mu \nu} = 
- \frac{\partial \theta_\perp}{\partial r} \cosh \theta_\perp 
- \frac{\partial \theta_\perp}{\partial \tau} \sinh \theta_\perp,
\label{thetaX}
\end{equation}
\begin{equation}
\theta_Y = Y_\mu Y_\nu \theta^{\mu \nu} = 
- \frac{\sinh \theta_\perp}{r},
\label{thetaY}
\end{equation}
and
\begin{equation}
\theta_Z = Z_\mu Z_\nu \theta^{\mu \nu} = 
- \frac{\cosh \theta_\perp}{\tau}.
\label{thetaZ}
\end{equation}

The contraction of the tensors $\Delta^{\mu \nu}$ and $\theta^{\mu \nu}$ gives the volume 
expansion parameter
\begin{equation}
\theta = \Delta^{\mu \nu} \theta_{\mu \nu}.
\label{theta1}
\end{equation}
Equations (\ref{Delta}) and (\ref{theta1}) yield
\begin{equation}
\theta = -\theta_X - \theta_Y - \theta_Z.
\label{theta2}
\end{equation}
Substituting Eqs. (\ref{thetaX})--(\ref{thetaZ}) in Eq. (\ref{theta2}) we find that this formula is
consistent with the definition $\theta = \partial_\mu U^\mu$.

The {\it shear} tensor $\sigma_{\mu \nu}$ is defined by the formula
\begin{equation}
\sigma_{\mu \nu} = \theta_{\mu \nu} - \frac{1}{3} \Delta_{\mu \nu} \theta.
\label{sigma1}
\end{equation}
With the help of the decompositions (\ref{Delta}) and (\ref{theta-dec}) we may write
\begin{equation}
\sigma^{\mu \nu} = \sigma_X X^\mu X^\nu + \sigma_Y Y^\mu Y^\nu +  \sigma_Z Z^\mu Z^\nu,
\label{sigma-dec}
\end{equation}
where
\begin{eqnarray}
\sigma_X &=& \frac{\theta}{3}+\theta_X = \frac{\cosh \theta_\perp}{3 \tau} + 
\frac{\sinh\theta_\perp}{3r} \label{sigmaX} \\
&& 
-\frac{2}{3} \frac{\partial\theta_\perp}{\partial \tau} \sinh\theta_\perp 
-\frac{2}{3} \frac{\partial\theta_\perp}{\partial r} \cosh\theta_\perp , \nonumber 
\end{eqnarray}
\begin{eqnarray}
\sigma_Y &=& \frac{\theta}{3}+\theta_Y = \frac{\cosh \theta_\perp}{3 \tau} -
\frac{2 \sinh\theta_\perp}{3r} \label{sigmaY} \\
&& 
+\frac{1}{3} \frac{\partial\theta_\perp}{\partial \tau} \sinh\theta_\perp 
+\frac{1}{3} \frac{\partial\theta_\perp}{\partial r} \cosh\theta_\perp , \nonumber 
\end{eqnarray}
and
\begin{eqnarray}
\sigma_Z &=& \frac{\theta}{3}+\theta_Z \label{sigmaZ} 
= -\frac{2\cosh \theta_\perp}{3 \tau} +
\frac{\sinh\theta_\perp}{3r} \\
&& 
+\frac{1}{3} \frac{\partial\theta_\perp}{\partial \tau} \sinh\theta_\perp 
+\frac{1}{3} \frac{\partial\theta_\perp}{\partial r} \cosh\theta_\perp . \nonumber 
\end{eqnarray}
In agreement with general requirements we find that 
\begin{equation}
\sigma_X+\sigma_Y+\sigma_Z=0.
\label{sum-sigma}
\end{equation}
In the case where the radial flow is absent \mbox{$\sigma_X = \sigma_Y = 1/(3 \tau)$} and \mbox{$\sigma_Z = -2/(3 \tau)$}, which agrees with earlier findings \cite{Muronga:2003ta}.

%%%%%%%%%%%%%%%%%%%%%%%%%%%%%%%%%%%%%%%%%%%%%%%%%%%%%%%%%%%%%%%%%%%%%%%%%%%%%%%%%%%%%%%%%%%%%%%%%%%%%
%%%%%%%%%%%%%%%%%%%%%%%%%%%%%%%%%%%%%%%%%%%%%%%%%%%%%%%%%%%%%%%%%%%%%%%%%%%%%%%%%%%%%%%%%%%%%%%%%%%%%
\section{Energy-momentum tensor}
\label{sect:Tmunu}
%%%%%%%%%%%%%%%%%%%%%%%%%%%%%%%%%%%%%%%%%%%%%%%%%%%%%%%%%%%%%%%%%%%%%%%%%%%%%%%%%%%%%%%%%%%%%%%%%%%%%
%%%%%%%%%%%%%%%%%%%%%%%%%%%%%%%%%%%%%%%%%%%%%%%%%%%%%%%%%%%%%%%%%%%%%%%%%%%%%%%%%%%%%%%%%%%%%%%%%%%%%

The energy-momentum tensors of the systems considered below in this paper have the following structure
\begin{eqnarray}
T^{\mu \nu} = \varepsilon U^\mu U^\nu + P_X X^\mu X^\nu + P_Y Y^\mu Y^\nu + P_Z Z^\mu Z^\nu. 
\label{enmomten}
\end{eqnarray}
The quantity $\varepsilon$ is the energy density, while $P_X, P_Y$ and $P_Z$ are three different pressure components. In LRF the energy-momentum tensor has the diagonal structure,
\begin{equation}
T^{\mu \nu} =  \left(
\begin{array}{cccc}
\varepsilon & 0 & 0 & 0 \\
0 & P_X & 0 & 0 \\
0 & 0 & P_Y & 0 \\
0 & 0 & 0 & P_Z
\end{array} \right).
\label{Tmunuarray}
\end{equation}
Since we consider boost-invariant and cylindrically symmetric systems, $\varepsilon, P_X, P_Y$ and $P_Z$ may depend only on the (longitudinal) proper time
\begin{equation}
\tau = \sqrt{t^2 - z^2}
\label{tau}
\end{equation} 
and radial distance
\begin{equation}
r = \sqrt{x^2 + y^2}.
\label{r}
\end{equation} 

The hydrodynamic equations include the energy-momentum conservation law
\begin{equation}
\partial_\mu T^{\mu \nu} = 0.
\label{enmomCL1}
\end{equation}
Using the form of the energy-momentum tensor (\ref{enmomten}) in (\ref{enmomCL1}) and projecting the result on $U_\nu$, $Z_\nu$, $X_\nu$ and $Y_\nu$ one gets four equations
\begin{eqnarray}
&& {\dot \varepsilon} + \varepsilon \partial_\mu U^\mu 
+  P_Z U_\nu Z^\mu \partial_\mu Z^\nu  \label{enmomCLU} \\
&& \quad + P_X U_\nu X^\mu \partial_\mu X^\nu + P_Y U_\nu Y^\mu \partial_\mu Y^\nu = 0, \nonumber \\
&& \varepsilon Z_\nu {\dot U}^\nu  - Z^\mu \partial_\mu P_Z 
-  P_Z \partial_\mu Z^\mu \label{enmomCLZ} \\
&& \quad + P_X Z_\nu X^\mu \partial_\mu X^\nu + P_Y Z_\nu Y^\mu \partial_\mu Y^\nu = 0, \nonumber \\
&& \varepsilon X_\nu {\dot U}^\nu  - X^\mu \partial_\mu P_X 
-  P_X \partial_\mu X^\mu \label{enmomCLX} \\
&& \quad + P_Y X_\nu Y^\mu \partial_\mu Y^\nu + P_Z X_\nu Z^\mu \partial_\mu Z^\nu = 0, \nonumber \\
&& \varepsilon Y_\nu {\dot U}^\nu  - Y^\mu \partial_\mu P_Y 
-  P_Y \partial_\mu Y^\mu \label{enmomCLY} \\
&& \quad + P_X  Y_\nu X^\mu \partial_\mu X^\nu + P_Z Y_\nu Z^\mu \partial_\mu Z^\nu = 0. \nonumber
\end{eqnarray}
Here the dot denotes the total time derivative (the operator $U^\alpha \partial_\alpha$). With the help of the relations 
\begin{eqnarray}
&& Z^\mu \partial_\mu = \frac{\partial}{\tau \partial \eta_\parallel}, \quad
\partial_\mu Z^\mu = 0, \quad {\dot Z}^\nu = 0, \nonumber \\
&& Y^\mu \partial_\mu = \frac{\partial}{r \partial \phi}, \quad
\partial_\mu Y^\mu = 0, \quad {\dot Y}^\nu = 0, \nonumber \\
&& X^\mu \partial_\mu Z^\nu = 0, \quad  X^\mu \partial_\mu Y^\nu = 0,
\end{eqnarray}
one can show that Eqs. (\ref{enmomCLZ}) and (\ref{enmomCLY}) are automatically fulfilled. In this way, we are left with only two independent equations:
\begin{eqnarray}
&& \left( \cosh \theta_\perp \frac{\partial}{\partial \tau} 
+ \sinh \theta_\perp \frac{\partial}{\partial r} \right) \varepsilon  \label{enmomCLU1} \\
&& + \varepsilon \left[ \cosh \theta_\perp \left( \frac{1}{\tau} 
+ \frac{\partial \theta_\perp}{\partial r} \right) + \sinh \theta_\perp \left( 
\frac{1}{r} + \frac{\partial \theta_\perp}{\partial \tau} \right) \right] \nonumber \\
&& + P_X \left( \cosh \theta_\perp \frac{\partial \theta_\perp }{\partial r} 
+ \sinh \theta_\perp \frac{\partial \theta_\perp }{\partial \tau} \right) \nonumber \\
&& P_Y \frac{\sinh \theta_\perp}{r} +P_Z \frac{\cosh \theta_\perp}{\tau} = 0 \nonumber
\end{eqnarray}
and
\begin{eqnarray}
&& \left( \sinh \theta_\perp \frac{\partial}{\partial \tau} 
+ \cosh \theta_\perp \frac{\partial}{\partial r} \right) P_X  \label{enmomCLX1} \\
&& + \varepsilon \left( \sinh \theta_\perp \frac{\partial \theta_\perp }{\partial r} 
+ \cosh \theta_\perp \frac{\partial \theta_\perp }{\partial \tau} \right)  \nonumber \\
&& + P_X \left[ \sinh \theta_\perp \left( \frac{1}{\tau} 
+ \frac{\partial \theta_\perp}{\partial r} \right) + \cosh \theta_\perp \left( 
\frac{1}{r} + \frac{\partial \theta_\perp}{\partial \tau} \right) \right] \nonumber \\
&& - P_Y \frac{\cosh \theta_\perp}{r} - P_Z \frac{\sinh \theta_\perp}{\tau} = 0. \nonumber
\end{eqnarray}

%%%%%%%%%%%%%%%%%%%%%%%%%%%%%%%%%%%%%%%%%%%%%%%%%%%%%%%%%%%%%%%%%%%%%%%%%%%%%%%%%%%%%%%%%%%%%%%%%%%%%
%%%%%%%%%%%%%%%%%%%%%%%%%%%%%%%%%%%%%%%%%%%%%%%%%%%%%%%%%%%%%%%%%%%%%%%%%%%%%%%%%%%%%%%%%%%%%%%%%%%%%
\section{Israel-Stewart theory}
\label{sect:df}
%%%%%%%%%%%%%%%%%%%%%%%%%%%%%%%%%%%%%%%%%%%%%%%%%%%%%%%%%%%%%%%%%%%%%%%%%%%%%%%%%%%%%%%%%%%%%%%%%%%%%
%%%%%%%%%%%%%%%%%%%%%%%%%%%%%%%%%%%%%%%%%%%%%%%%%%%%%%%%%%%%%%%%%%%%%%%%%%%%%%%%%%%%%%%%%%%%%%%%%%%%%

%%%%%%%%%%%%%%%%%%%%%%%%%%%%%%%%%%%%%%%%%%%%%%%%%%%%%%%%%%%%%%%%%%%%%%%%%%%%%%%%%%%%%%%%%%%%%%%%%%%%%
%%%%%%%%%%%%%%%%%%%%%%%%%%%%%%%%%%%%%%%%%%%%%%%%%%%%%%%%%%%%%%%%%%%%%%%%%%%%%%%%%%%%%%%%%%%%%%%%%%%%%
\subsection{Stress tensor}
\label{sect:stress}
%%%%%%%%%%%%%%%%%%%%%%%%%%%%%%%%%%%%%%%%%%%%%%%%%%%%%%%%%%%%%%%%%%%%%%%%%%%%%%%%%%%%%%%%%%%%%%%%%%%%%
%%%%%%%%%%%%%%%%%%%%%%%%%%%%%%%%%%%%%%%%%%%%%%%%%%%%%%%%%%%%%%%%%%%%%%%%%%%%%%%%%%%%%%%%%%%%%%%%%%%%%

In the Israel-Stewart theory , the crucial role is played by the stress tensor $\pi^{\mu \nu}$ that satisfies the following differential equation \cite{Israel:1979wp,Muronga:2003ta}
\begin{equation}
\tau_\pi \Delta^\alpha_\mu \Delta^\beta_\nu {\dot \pi}_{\alpha \beta} + \pi_{\mu \nu}
= 2 \eta \sigma_{\mu \nu} + F_\eta \pi_{\mu \nu}.
\label{pi-eq-1}
\end{equation}
Here $\tau_\pi$ is the relaxation time, $\eta$ is the shear viscosity,  and $F_\eta$ is our abbreviation for the scalar quantity
\begin{equation}
F_\eta = -\eta T \partial_\lambda \left( \frac{\alpha_1}{T} U^\lambda \right),
\label{Feta}
\end{equation}
where $T$ is the temperature and $\alpha_1$ is one of the kinetic coefficients appearing in the Israel-Stewart theory.

The structure of the shear tensor, Eq. (\ref{sigma-dec}), suggests that we may seek the solutions of Eq. (\ref{pi-eq-1}) in the form analogous to Eqs. (\ref{theta-dec}) and (\ref{sigma-dec}), namely
\begin{equation}
\pi^{\mu \nu} = \pi_X X^\mu X^\nu + \pi_Y Y^\mu Y^\nu +  \pi_Z Z^\mu Z^\nu.
\label{pi-dec}
\end{equation}
The condition $\Delta^{\mu \nu} \pi_{\mu \nu} =0$ leads to the constraint
\begin{equation}
\pi_X + \pi_Y + \pi_Z = 0.
\label{sum-pi}
\end{equation}

The time derivative of $\pi^{\mu \nu}$ generates nine terms. Since $X^\mu, Y^\mu$ and $Z^\mu$ are the eigenvectors of the projection operator $\Delta^{\mu \nu}$, see Eq. (\ref{eigen}), we find
\begin{eqnarray}
&& \Delta^\alpha_\mu \Delta^\beta_\nu  {\dot \pi}_{\alpha \beta} = \nonumber \\
&&  \,\,\, {\dot \pi}_X X_\mu X_\nu +  \pi_X \Delta^\alpha_\mu {\dot X}_\alpha X_\nu 
+ \pi_X X_\mu \Delta^\beta_\nu {\dot X}_\beta
\nonumber \\
&& + {\dot \pi}_Y Y_\mu Y_\nu + \pi_Y \Delta^\alpha_\mu {\dot Y}_\alpha Y_\nu
+  \pi_Y Y_\mu \Delta^\beta_\nu {\dot Y}_\beta
\nonumber \\
&& + {\dot \pi}_Z Z_\mu Z_\nu + \pi_Z \Delta^\alpha_\mu {\dot Z}_\alpha Z_\nu 
+  \pi_Z Z_\mu \Delta^\beta_\nu {\dot Z}_\beta.
\nonumber \\
\label{sigma-dec-1}
\end{eqnarray}
For boost-invariant and cylindrically symmetric systems, the explicit calculations show that the terms $\Delta^\alpha_\mu {\dot X}_\alpha$, $\Delta^\alpha_\mu {\dot Y}_\alpha$, and $\Delta^\alpha_\mu {\dot Z}_\alpha$ vanish. Therefore, the decomposition (\ref{pi-dec}) is indeed appropriate, and Eq. (\ref{pi-eq-1}) splits into three scalar equations
\begin{eqnarray}
\tau_\pi {\dot \pi}_X + \pi_X &=& 2 \eta \sigma_X + F_\eta \pi_X, \nonumber \\
\tau_\pi {\dot \pi}_Y + \pi_Y &=& 2 \eta \sigma_Y + F_\eta \pi_Y, \nonumber \\
\tau_\pi {\dot \pi}_Z + \pi_Z &=& 2 \eta \sigma_Z + F_\eta \pi_Z.
\label{pi-eq-2}
\end{eqnarray}
Due to the constraints (\ref{sum-sigma}) and (\ref{sum-pi}), only two equations in (\ref{pi-eq-2}) are independent.

%%%%%%%%%%%%%%%%%%%%%%%%%%%%%%%%%%%%%%%%%%%%%%%%%%%%%%%%%%%%%%%%%%%%%%%%%%%%%%%%%%%%%%%%%%%%%%%%%%%%%
%%%%%%%%%%%%%%%%%%%%%%%%%%%%%%%%%%%%%%%%%%%%%%%%%%%%%%%%%%%%%%%%%%%%%%%%%%%%%%%%%%%%%%%%%%%%%%%%%%%%%
\subsection{Bulk viscosity}
\label{sect:bulk}
%%%%%%%%%%%%%%%%%%%%%%%%%%%%%%%%%%%%%%%%%%%%%%%%%%%%%%%%%%%%%%%%%%%%%%%%%%%%%%%%%%%%%%%%%%%%%%%%%%%%%
%%%%%%%%%%%%%%%%%%%%%%%%%%%%%%%%%%%%%%%%%%%%%%%%%%%%%%%%%%%%%%%%%%%%%%%%%%%%%%%%%%%%%%%%%%%%%%%%%%%%%

The isotropic correction to pressure, $\Pi$, satisfies the equation
\begin{equation}
\tau_\Pi {\dot \Pi} + \Pi = -\zeta \theta + F_\zeta \Pi,
\label{Pi-eq-1}
\end{equation} 
where $\tau_\Pi$ is the relaxation time for $\Pi$, $\zeta$ is the bulk viscosity, and
\begin{equation}
F_\zeta =  -\frac{1}{2} \zeta T \partial_\lambda \left( \frac{\alpha_0}{T} U^\lambda \right),
\label{Fzeta}
\end{equation}
where $\alpha_0$ is another kinetic coefficient appearing in the Israel-Stewart theory.

%%%%%%%%%%%%%%%%%%%%%%%%%%%%%%%%%%%%%%%%%%%%%%%%%%%%%%%%%%%%%%%%%%%%%%%%%%%%%%%%%%%%%%%%%%%%%%%%%%%%%
%%%%%%%%%%%%%%%%%%%%%%%%%%%%%%%%%%%%%%%%%%%%%%%%%%%%%%%%%%%%%%%%%%%%%%%%%%%%%%%%%%%%%%%%%%%%%%%%%%%%%
\subsection{Energy-momentum tensor and hydrodynamic equations}
\label{sect:enmomtenIS}
%%%%%%%%%%%%%%%%%%%%%%%%%%%%%%%%%%%%%%%%%%%%%%%%%%%%%%%%%%%%%%%%%%%%%%%%%%%%%%%%%%%%%%%%%%%%%%%%%%%%%
%%%%%%%%%%%%%%%%%%%%%%%%%%%%%%%%%%%%%%%%%%%%%%%%%%%%%%%%%%%%%%%%%%%%%%%%%%%%%%%%%%%%%%%%%%%%%%%%%%%%%

In the Israel-Stewart theory, the energy momentum tensor has the following form
\begin{equation}
T^{\mu\nu} = \varepsilon U^\mu U^\nu - P_{\rm eq} \Delta^{\mu \nu} + \pi^{\mu \nu} -\Pi \Delta^{\mu \nu},
\label{enmomtenIS}
\end{equation}
where $P_{\rm eq} $ is the equilibrium pressure connected with the energy density by the equation of state, $P_{\rm eq}=P_{\rm eq}(\varepsilon)$, and $\Pi$ is the isotropic correction to pressure. A simple comparison of Eqs. (\ref{enmomtenIS}) and (\ref{enmomten}) leads to the identifications
\begin{eqnarray}
P_X &=& P_{\rm eq}  + \Pi + \pi_X, P_Y = P_{\rm eq}  + \Pi + \pi_Y, \nonumber \\
P_Z &=& P_{\rm eq}  + \Pi + \pi_Z = P_{\rm eq}  + \Pi - \pi_X - \pi_Y.
\label{PXYZ_IS}
\end{eqnarray}

Substituting Eqs. (\ref{PXYZ_IS}) into Eqs. (\ref{enmomCLU1}) and (\ref{enmomCLX1}) we obtain two equations for five unknown functions: $\varepsilon$, $\theta_\perp$, $\Pi$, $\pi_X$, and $\pi_Y$. The two first equations in (\ref{pi-eq-2}) as well as Eq. (\ref{Pi-eq-1}) should be included as the three extra equations needed to close this system.

%%%%%%%%%%%%%%%%%%%%%%%%%%%%%%%%%%%%%%%%%%%%%%%%%%%%%%%%%%%%%%%%%%%%%%%%%%%%%%%%%%%%%%%%%%%%%%%%%%%%%
%%%%%%%%%%%%%%%%%%%%%%%%%%%%%%%%%%%%%%%%%%%%%%%%%%%%%%%%%%%%%%%%%%%%%%%%%%%%%%%%%%%%%%%%%%%%%%%%%%%%%
\section{ADHYDRO model}
\label{sect:adhydro}
%%%%%%%%%%%%%%%%%%%%%%%%%%%%%%%%%%%%%%%%%%%%%%%%%%%%%%%%%%%%%%%%%%%%%%%%%%%%%%%%%%%%%%%%%%%%%%%%%%%%%
%%%%%%%%%%%%%%%%%%%%%%%%%%%%%%%%%%%%%%%%%%%%%%%%%%%%%%%%%%%%%%%%%%%%%%%%%%%%%%%%%%%%%%%%%%%%%%%%%%%%%

The Israel-Stewart theory describes the system that is close to local equilibrium. Formally, this means that the corrections to pressure (the quantities $\Pi$, $\pi_X$, and $\pi_Y$) should be small compared to $P_{\rm eq}$. This condition cannot be fulfilled at the very early stages of heavy ion collisions. In the limit $\tau \to 0$ the components of the shear tensor, see Eqs.  (\ref{sigmaX})--(\ref{sigmaZ}), diverge and induce very large changes of $\pi_X$, $\pi_Y$, and $\pi_Z$ through Eqs. (\ref{pi-eq-2}). This leads to strong deviations from local equilibrium. 

In this situation, one tries to construct phenomenological models of the very early stages which grasp the essential features of the produced matter and may describe effectively the early dynamics. On the basis of microscopic models of heavy-ion collisions, we expect that the system formed at the very early stages  of heavy-ion collisions is highly anisotropic --- the two transverse pressures are equal and much larger than the longitudinal pressure. 

Such anisotropic systems are described most often by the anisotropic distribution functions which have the form of the squeezed or stretched Boltzmann equilibrium distributions for massless partons (this is often called the Romatschke-Strickland ansatz \cite{Romatschke:2003ms})
\begin{eqnarray}
f &=& g \exp\left[ -\frac{1}{\lambda_\perp}\sqrt{(p \cdot U)^2 + (x-1) (p \cdot Z)^2}\right].
\label{f}
\end{eqnarray}
Here $p$ is the particle's four momentum, $g$ denotes the number of the internal degrees of freedom, $\lambda_\perp$ may be interpreted as the temperature of the transverse degrees of freedom, and $x$ is the anisotropy parameter. The energy-momentum tensor of the system described by the distribution function (\ref{f}) has the form \cite{Florkowski:2010cf,Ryblewski:2010bs,Ryblewski:2010ch,Ryblewski:2011aq}~\footnote{In the original papers the four-vector $Z^\mu$ is denoted as $V^\mu$.}
\begin{equation}
T^{\mu \nu} = \left( \varepsilon  + P_\perp\right) U^{\mu}U^{\nu} - P_\perp \, g^{\mu\nu} - (P_\perp - P_\parallel) Z^{\mu}Z^{\nu}. 
\label{TmunuA1}
\end{equation}
This form agrees with Eq. (\ref{enmomten}) if we set
\begin{equation}
P_X = P_Y = P_\perp, \quad P_Z = P_\parallel.
\label{PXYZ_ADHYDRO}
\end{equation}

In the case described by the distribution function (\ref{f}), the energy density and the two pressures may be expressed as functions of the non-equilibrium entropy density $\sigma$ and the anisotropy parameter $x$ \cite{Ryblewski:2011aq},
\begin{eqnarray}
\varepsilon(\sigma,x) &=&  \varepsilon_{\rm eq}(\sigma) r(x), \label{epsilon2a}  \\ \nonumber \\
P_\perp(\sigma,x) &=&  P_{\rm eq}(\sigma) \left[r(x) + 3 x r^\prime(x) \right],  \label{PT2a}  \\ \nonumber \\
P_\parallel(\sigma,x) &=&  P_{\rm eq}(\sigma) \left[r(x) - 6 x r^\prime(x) \right]. \label{PL2a} 
\end{eqnarray}
We emphasize that $\varepsilon_{\rm eq}(\sigma)$ and $P_{\rm eq}(\sigma)$ are equilibrium expressions for the energy density and pressure  \cite{Florkowski:2010zz} but the argument is a non-equilibrium value of the entropy density 
\begin{equation}
\varepsilon_{\rm eq}(\sigma) = 3 P_{\rm eq}(\sigma) = \frac{3g}{\pi^2} \left( \frac{\pi^2 \sigma}{4g} \right)^{4/3}.
\end{equation}
The function $r(x)$ has the form
\begin{eqnarray}
r(x) = \frac{x^{-\frac{1}{3}}}{2} \left[ 1 + \frac{x \arctan\sqrt{x-1}}{\sqrt{x-1}}\right].
\label{rx}  \\ \nonumber
\end{eqnarray} 

Substituting Eqs. (\ref{PXYZ_ADHYDRO})--(\ref{rx}) into Eqs. (\ref{enmomCLU1}) and (\ref{enmomCLX1}) (which follow directly from the energy-momentum conservation law) we obtain two equations for three unknown functions: $\sigma$, $x$, and $\theta_\perp$. The third equation follows from the ansatz describing the entropy production in the system,
\begin{eqnarray}
\partial_\mu \sigma^{\mu} &=& \Sigma(\sigma,x). \label{engrow}
\end{eqnarray}
The entropy source $\Sigma$ is taken in the form
\begin{equation}
\Sigma(\sigma,x) = \frac{(1-\sqrt{x})^2}{\sqrt{x}} \frac{\sigma}{\tau_{\rm eq}}.
\label{ansatz1}
\end{equation} 
The quantity ${\tau_{\rm eq}}$ is a timescale parameter. The form (\ref{ansatz1}) guarantees that $\Sigma \geq 0$ and $\Sigma(\sigma,x=1)=0$. We stress that ADHYDRO accepts other reasonable definitions of the entropy source. In particular, it would be interesting in this context to use the forms motivated by the AdS/CFT correspondence \cite{Heller:2011ju}.

%%%%%%%%%%%%%%%%%%%%%%%%%%%%%%%%%%%%%%%%%%%%%%%%%%%%%%%%%%%%%%%%%%%%%%%%%%%%%%%%%%%%%%%%%%%%%%%%%%%%%
%%%%%%%%%%%%%%%%%%%%%%%%%%%%%%%%%%%%%%%%%%%%%%%%%%%%%%%%%%%%%%%%%%%%%%%%%%%%%%%%%%%%%%%%%%%%%%%%%%%%%
\section{Matching conditions between ADHYDRO model and Israel-Stewart theory}
\label{sect:adhydro}
%%%%%%%%%%%%%%%%%%%%%%%%%%%%%%%%%%%%%%%%%%%%%%%%%%%%%%%%%%%%%%%%%%%%%%%%%%%%%%%%%%%%%%%%%%%%%%%%%%%%%
%%%%%%%%%%%%%%%%%%%%%%%%%%%%%%%%%%%%%%%%%%%%%%%%%%%%%%%%%%%%%%%%%%%%%%%%%%%%%%%%%%%%%%%%%%%%%%%%%%%%%

In this Section, we show how the initial evolution of the system described by the ADHYDRO model may be matched to a later non-equilibrium evolution governed by the Israel-Stewart theory. 

We propose to do the matching at the transition proper time $\tau_{\rm tr}$ when the anisotropy parameter $x$ becomes close to unity in the whole space, i.e., when the condition $|x(\tau=\tau_{\rm tr},r)-1| \ll 1$ is satisfied for all values of $r$. Our earlier calculations done within the ADHYDRO framework \cite{Ryblewski:2011aq} show that if the initial value of $x$ is independent of $r$, the later values of $x$ depend weakly on $r$, hence, it makes sense to use the value of the transition time that is to large extent independent of $r$. Certainly, one should always check the sensitivity of the obtained results with respect to the chosen value of $\tau_{\rm tr}$. The acceptable results should exhibit weak dependence on $\tau_{\rm tr}$.

We emphasize that the proposed matching procedure differs from our previous strategy where the ADHYDRO model was used to describe the whole evolution of the system; from a highly-anisotropic initial stage to hadronic freeze-out \cite{Florkowski:2010cf,Ryblewski:2010bs,Ryblewski:2010ch,Ryblewski:2011aq}. The use of the ADHYDRO model {\it alone} implies a smooth switching from a highly-anisotropic phase (where $x \gg 1$ or $x \ll 1$) to the phase described by the perfect-fluid hydrodynamics (where $x \approx 1$). Moreover, as it has been shown in \cite{Florkowski:2010cf,Martinez:2010sc}, for purely longitudinal and boost-invariant expansion of matter, ADHYDRO agrees with the Israel-Stewart framework in the intermediate region where $|x-1| \ll 1$. On the other hand, if the transverse expansion is included, the presence of non-negligible shear viscosity triggers differences between the two components of the transverse pressure, $P_X$ and $P_Y$. This effect is not included in the ADHYDRO model. Therefore, in the approaches that include transverse expansion and noticeable effects of viscosity, the matching proposed below is in our opinion more appropriate than the use of ADHYDRO alone. However, the use of the ADHYDRO model is reasonable for the situations where the effects of viscosity at the later stages of the collisions may be neglected. 

%%%%%%%%%%%%%%%%%%%%%%%%%%%%%%%%%%%%%%%%%%%%%%%%%%%%%%%%%%%%%%%%%%%%%%%%%%%%%%%%%%%%%%%%%%%%%%%%%%%%%
%%%%%%%%%%%%%%%%%%%%%%%%%%%%%%%%%%%%%%%%%%%%%%%%%%%%%%%%%%%%%%%%%%%%%%%%%%%%%%%%%%%%%%%%%%%%%%%%%%%%%
\subsection{Energy-momentum matching}
\label{sect:enmom-match}
%%%%%%%%%%%%%%%%%%%%%%%%%%%%%%%%%%%%%%%%%%%%%%%%%%%%%%%%%%%%%%%%%%%%%%%%%%%%%%%%%%%%%%%%%%%%%%%%%%%%%
%%%%%%%%%%%%%%%%%%%%%%%%%%%%%%%%%%%%%%%%%%%%%%%%%%%%%%%%%%%%%%%%%%%%%%%%%%%%%%%%%%%%%%%%%%%%%%%%%%%%%

In order to connect the solutions of the ADHYDRO model with the solutions of the Israel-Stewart theory, we demand that
all thermodynamics- and hydrodynamics-like quantities are continuous across the transition boundary fixed at the transition time $\tau_{\rm tr}$, namely:

i) the energy density $\varepsilon$ is the same on both sides of the transition and identified with the equilibrium energy density in the Israel-Stewart theory, $\varepsilon = \varepsilon_{\rm eq}$,

ii)  the transverse flow, quantified by the value of $\theta_\perp$, is the same at the end of the ADHYDRO stage and at the beginning of the Israel-Stewart stage,

iii) also the three components of pressure are the same, namely
\begin{eqnarray}
P_{\rm eq} + \Pi + \pi_X &=& P_\perp, \nonumber \\  
P_{\rm eq} + \Pi + \pi_Y &=& P_\perp, \label{pressure-match} \\  
P_{\rm eq} + \Pi + \pi_Z &=& P_\parallel. \nonumber
\end{eqnarray}
Here the values on the right-hand-side are obtained at the end of the ADHYDRO evolution, at $\tau=\tau_{\rm tr}$, and treated as the input for the stage described by the Israel-Stewart equations for $\tau \geq \tau_{\rm tr}$. From (\ref{pressure-match}) we find first that $\pi_Y = \pi_X$ and $\pi_Z = -2 \pi_X$. In the next step we find that
\begin{equation}
\pi_X = \frac{P_\perp-P_\parallel}{3} = 
\varepsilon_{\rm eq}(\sigma) x r'(x)
\approx  \varepsilon_{\rm eq}(\sigma_{\rm eq}) \frac{4(x-1)}{45}
\label{pi-X-match}
\end{equation}
and
\begin{equation}
P_{\rm eq} + \Pi = \frac{2 P_\perp + P_\parallel}{3} = \frac{\varepsilon}{3} = \frac{\varepsilon_{\rm eq}}{3}.
\label{Peq+pi}
\end{equation}
For massless particles considered here, $P_{\rm eq} = \varepsilon_{\rm eq}/3$ which implies $\Pi=0$.

Finally, we demand that iv) the entropy density is the same before and after the transition. This is discussed in more detail in the next Section.

%%%%%%%%%%%%%%%%%%%%%%%%%%%%%%%%%%%%%%%%%%%%%%%%%%%%%%%%%%%%%%%%%%%%%%%%%%%%%%%%%%%%%%%%%%%%%%%%%%%%%
%%%%%%%%%%%%%%%%%%%%%%%%%%%%%%%%%%%%%%%%%%%%%%%%%%%%%%%%%%%%%%%%%%%%%%%%%%%%%%%%%%%%%%%%%%%%%%%%%%%%%
\subsection{Entropy matching}
\label{sect:entropy-match}
%%%%%%%%%%%%%%%%%%%%%%%%%%%%%%%%%%%%%%%%%%%%%%%%%%%%%%%%%%%%%%%%%%%%%%%%%%%%%%%%%%%%%%%%%%%%%%%%%%%%%
%%%%%%%%%%%%%%%%%%%%%%%%%%%%%%%%%%%%%%%%%%%%%%%%%%%%%%%%%%%%%%%%%%%%%%%%%%%%%%%%%%%%%%%%%%%%%%%%%%%%%

It is interesting to see in more detail how the last condition is realized in practice. First, we start with the ADHYDRO formulation. If the energy density $\varepsilon$ corresponds to the equilibrium energy density, see the condition i), the corresponding equilibrium entropy density may be obtained with the help of the inverse function to the function $\varepsilon_{\rm eq}(\sigma)$. In this way, expanding the function $r(x)$ at $x=1$, we find
\begin{eqnarray}
\sigma_{\rm eq} &=& \varepsilon_{\rm eq}^{-1}\left[ \varepsilon_{\rm eq}(\sigma) r(x) \right] \nonumber \\
&\approx & \varepsilon_{\rm eq}^{-1}\left[ \varepsilon_{\rm eq}(\sigma) + \varepsilon_{\rm eq}(\sigma) \frac{2 (x-1)^2}{45} \right] \label{entr-match-1} \\
&\approx & \sigma + [ d\varepsilon_{\rm eq}(\sigma)/d\sigma]^{-1} \varepsilon_{\rm eq}(\sigma) \frac{2 (x-1)^2}{45}. \nonumber 
\end{eqnarray}
In the leading order in deviations from the equilibrium,  we may replace $\sigma$ by $\sigma_{\rm eq}$ in the second term in the last line of (\ref{entr-match-1}). Using the thermodynamic identity for the system of massless particles we find
\begin{equation}
[d\varepsilon_{\rm eq}(\sigma_{\rm eq})/d\sigma_{\rm eq}]^{-1} \varepsilon_{\rm eq}(\sigma_{\rm eq}) 
=  \frac{\varepsilon_{\rm eq}(\sigma_{\rm eq}) }{T} = \frac{3}{4} \sigma_{\rm eq}.
\end{equation}
Combing the last two results we find
\begin{equation}
\sigma = \sigma_{\rm eq} \left(1 - \frac{(x-1)^2}{30}  \right).
\label{sigma-adhydro}
\end{equation}

Now we calculate the connection between the equilibrium and non-equilibrium entropy density using the Israel-Stewart theory. The basic relation in this context has the form \cite{Muronga:2003ta}
\begin{equation}
\sigma = \sigma_{\rm eq} - \beta_0 \frac{\Pi^2}{2T}
-\beta_2 \frac{ \pi_{\lambda \nu} \pi^{\lambda \nu} }{2T}.
\end{equation}
In our case we have $\beta_0 = \tau_\Pi/\zeta$, $\beta_2 = \tau_\pi/(2\eta)$, and $\pi_{\lambda \nu} \pi^{\lambda \nu}=6\pi_X^2$. We also use the result 
\begin{equation}
\tau_\pi = \frac{5\eta}{T \sigma_{\rm eq}},
\label{taupiMS}
\end{equation}
which has been derived in Ref. \cite{Martinez:2009ry}. Following \cite{Martinez:2009ry} we note that expansion of anisotropic distributions around the equilibrium backgrounds does not lead to the situation described by the 14 Grad's ansatz, hence (\ref{taupiMS}) differs from the standard result by a factor of $6/5$.

The relations listed above allow us to write 
\begin{equation}
\sigma = \sigma_{\rm eq} \left(1 -  \frac{135 \, \pi_X^2}{32 \, \varepsilon_{\rm eq}^2} \right)
-  \frac{\tau_\Pi \Pi^2}{2\zeta T}.
\end{equation}
Using Eq. (\ref{pi-X-match}) in this equation we find
\begin{equation}
\sigma = \sigma_{\rm eq} \left(1 -  \frac{(x-1)^2}{30}  \right)
-  \frac{\tau_\Pi \Pi^2}{2\zeta T}.
\label{sigma-IS}
\end{equation}
Thus, we conclude that matching between ADHYDRO and the Israel-Stewart theory is continuous if the condition (\ref{taupiMS}) is fulfilled and $\Pi=0$ at the transition time (in agreement with our remarks following Eq. (\ref{Peq+pi})). 

We note that the condition (\ref{taupiMS}) guarantees that the entropy production in ADHYDRO has the same form as in the Israel-Stewart theory, which has been already shown in Refs. \cite{Florkowski:2010cf,Martinez:2010sc}. We also note that for $\tau \geq \tau_{\rm tr}$ , the two transverse pressure start to differ from each other,  since their dynamics is governed by different components of the shear tensor.

%%%%%%%%%%%%%%%%%%%%%%%%%%%%%%%%%%%%%%%%%%%%%%%%%%%%%%%%%%%%%%%%%%%%%%%%%%%%%%%%%%%%%%%%%%%%%%%%%%%%
\section{Conclusions}
\label{sect:con}
%%%%%%%%%%%%%%%%%%%%%%%%%%%%%%%%%%%%%%%%%%%%%%%%%%%%%%%%%%%%%%%%%%%%%%%%%%%%%%%%%%%%%%%%%%%%%%%%%%%%%

In this paper we have introduced a basis of projection operators which allows for simple analysis
of dissipative fluid dynamics of boost-invariant and cylindrically symmetric systems. We have 
used this basis to analyze the equations of the Israel-Stewart theory and the ADHYDRO model. 
We have shown how the very early evolution of matter produced in heavy-ion collisions may be
described by the ADHYDRO equations combined with a later Israel-Stewart dynamics.

\end{document}